\documentclass[conference]{IEEEtran}
\IEEEoverridecommandlockouts
\usepackage{cite}
\usepackage{amsmath,amssymb,amsfonts}
\usepackage{algorithm}
\usepackage{algpseudocode}
\usepackage{graphicx}
\usepackage{textcomp}
\usepackage{xcolor}
\usepackage{multirow}
\usepackage{booktabs}
\usepackage{setspace}
\usepackage{comment}
\usepackage{enumitem}

\usepackage{subcaption}
\newlist{inlineroman}{enumerate*}{1}
\setlist[inlineroman]{itemjoin*={{ }},afterlabel=~,label=\roman*.}

\renewcommand{\baselinestretch}{0.97}

\algblock{Input}{EndInput}
\algnotext{EndInput}
\algblock{Output}{EndOutput}
\algnotext{EndOutput}

\def\BibTeX{{\rm B\kern-.05em{\sc i\kern-.025em b}\kern-.08em
    T\kern-.1667em\lower.7ex\hbox{E}\kern-.125emX}}

\begin{document}
\captionsetup[table]{labelfont=bf,textfont=normalfont}

\title{SpecReuse: Spectral Graph Reuse for Efficient Vision GNN Inference on FPGAs}
\author{
\IEEEauthorblockN{Isabella Bernhardt Eiliya, Anvitha Ramachandran, Dhruv Parikh, Viktor Prasanna}
\IEEEauthorblockA{
   University of Southern California, Los Angeles, California, USA \\
    \{ibernhar, alramach, dhruvash, prasanna\}@usc.edu}
}

\maketitle

\begin{abstract}
Dynamic Image Graph Construction (DIGC) is the primary performance bottleneck in FPGA acceleration of Vision Graph Neural Networks (ViGs), reconstructing graph connectivity at every layer through irregular, memory-intensive computation. Existing FPGA accelerators optimize DIGC but still execute it unconditionally, making repeated graph reconstruction a persistent source of latency and energy consumption. We propose the SpecReuse algorithm, which computes compact spectral descriptors of intermediate features and reuses previously constructed graphs when descriptor drift remains below a calibrated threshold. We further present the SpecReuse accelerator, an FPGA architecture that realizes graph reuse through lightweight hardware for spectral descriptor extraction and reuse control while remaining compatible with existing graph-construction accelerators. Experimental results demonstrate up to a $2.69\times$ speedup in end-to-end inference and approximately 58--65\% lower energy per inference with negligible FPGA resource overhead and minimal loss in classification accuracy.
\end{abstract}


%--------------------------------------------------------------
% add keywords for FPGA acceleration
%--------------------------------------------------------------

\section{Introduction}

Dynamic Image Graph Construction (DIGC) is the dominant performance bottleneck in FPGA acceleration of Vision Graph Neural Networks (ViGs). At each graph layer, the DIGC stage reconstructs graph connectivity from intermediate features through pairwise similarity computation, neighbor selection, and graph materialization, resulting in irregular computation, substantial memory traffic, and repeated execution throughout inference \cite{VisionGNN,han2023vihgnn,wu2023pvg,mobilevig2023,greedyvig2024,clustervig2025,wang2019dgcnn,li2019deepgcns}. Recent FPGA accelerators reduce the cost of DIGC through specialized dataflows, memory hierarchies, and pipeline optimizations \cite{yan2020hygcn,hu2021graphlily,zhang2023graphagile,digcfpga2025,graphleap2026}. However, these architectures continue to invoke the DIGC stage at every graph layer, making repeated graph construction a persistent source of latency and energy consumption despite increasingly efficient implementations.

This exposes an architectural opportunity that existing accelerators do not exploit: the DIGC stage executes unconditionally even when graph connectivity changes little between adjacent layers. Neighboring ViG layers frequently preserve similar graph connectivity, indicating that many DIGC executions perform redundant work. Rather than further accelerating DIGC, these observations motivate graph reuse as a complementary architectural optimization that reduces the frequency with which the DIGC stage executes.

SpecReuse realizes graph reuse as a lightweight prediction front-end that augments the graph construction pipeline ahead of the DIGC stage. The front-end computes compact spectral descriptors of intermediate features and uses spectral similarity to determine whether the previously constructed graph can be safely reused, leveraging spectral representations that effectively capture graph structure \cite{chung1997spectral,munir2025searchvig}. When graph reuse is selected, the DIGC stage is bypassed, eliminating unnecessary distance computation, sorting, and top-$k$ neighbor selection while remaining compatible with existing FPGA graph-construction accelerators \cite{yan2020hygcn,hu2021graphlily,zhang2023graphagile,digcfpga2025,graphleap2026}. Consequently, SpecReuse changes \emph{when} the DIGC stage executes rather than \emph{how} it is implemented, enabling lower inference latency and energy consumption with negligible additional hardware resources. The primary contributions of this work are:

\begin{itemize}[leftmargin=*]

\item We identify repeated DIGC execution as an architectural inefficiency in Vision GNN accelerators and characterize graph reuse opportunities across adjacent Vision GNN layers.

\item We propose SpecReuse, a lightweight spectral prediction front-end that augments the graph construction pipeline and enables graph reuse through spectral similarity estimation while remaining compatible with existing FPGA graph-construction accelerators.

\item We implement and evaluate SpecReuse on an AMD Alveo U280 FPGA across multiple Vision GNN backbones, achieving up to a $2.69\times$ end-to-end inference speedup and approximately $58$--$65\%$ lower energy per inference with only $1.07\%$ LUT, $0.56\%$ FF, $0.12\%$ DSP, and $0.88\%$ BRAM overhead. Offline threshold calibration enables up to $30.1\%$ graph reuse while limiting Top-1 accuracy degradation to at most $0.54$ percentage points.
\end{itemize}

%%%%%%%%%%%%%%%%%%%%%%%%%%%%%%%%%%%%%%%%%%%%%%%%%%%%%%%%%%%%%%%%%%%%%%%%%%%%%%%%
\section{Related Work}
\label{sec:related}

Vision Graph Neural Networks (ViGs) reconstruct graph connectivity
dynamically throughout inference to adapt neighborhood structure as feature
representations evolve
\cite{VisionGNN,han2023vihgnn,wu2023pvg,mobilevig2023}. Recent ViG
architectures improve graph quality and efficiency through progressive graph
construction, sparse neighborhood selection, hierarchical graph
representations, and structured graph formation
\cite{greedyvig2024,clustervig2025,DVHGNN}. WiGNet instead partitions the
image into local windows and restricts graph construction to each window,
reducing the cost of graph construction by shrinking the candidate
neighborhood rather than changing how often the graph is rebuilt
\cite{Spadaro_2025_WACV}. The repeated reconstruction of
graph connectivity is also central to dynamic graph learning architectures
such as Dynamic Graph CNNs and DeepGCNs, where graph topology evolves with the
underlying feature representation
\cite{wang2019dgcnn,li2019deepgcns,gilmer2017neural}. Collectively, these
approaches reduce graph-construction cost or improve graph quality, yet
continue to reconstruct graph connectivity throughout inference.

FPGA graph accelerators improve graph processing through customized
dataflows, specialized memory hierarchies, hardware pipelines, workload
balancing, and automated accelerator generation
\cite{yan2020hygcn,hu2021graphlily,zhang2023graphagile,awbgcn2020,boostgcn2021,gnnbuilder2023}.
Other accelerators target adaptability and latency directly: FP-GNN adjusts
precision and layer execution at runtime to accommodate diverse GNN workloads
\cite{Tian_2022_FP_GNN}, while LL-GNN targets low-latency inference for
high-energy-physics triggers through a streaming FPGA pipeline
\cite{Que_2024_LL_GNN}. These designs improve how graph convolution is
executed on FPGA but, like the accelerators above, operate on a graph that is
assumed to already be available.

A separate line of work accelerates the construction of the graph itself.
FNNG builds a dedicated FPGA pipeline for $k$-nearest-neighbor graph
construction \cite{Liu_2023_FNNG}, and L-FNNG extends this design to
large-scale graphs on a CPU-FPGA heterogeneous platform
\cite{He_2024_LFNNG}. Outside the vision domain, real-time graph-building
pipelines have also been proposed for particle-physics trigger applications,
where graph connectivity must likewise be rebuilt from streaming input data
\cite{Neu_2024_GraphGen}. These accelerators reduce the latency of a single
graph-construction pass, an optimization that is complementary to and could
be combined with the SpecReuse algorithm, which instead reduces how often a
graph-construction pass is issued in the first place. More recent architectures extend graph-construction acceleration to Vision
GNN inference specifically, accelerating the Dynamic Image Graph Construction
(DIGC) stage and overlapping graph construction with graph convolution
\cite{digcfpga2025,graphleap2026}. Although these accelerators substantially
reduce the cost of DIGC, they continue to execute the DIGC stage
unconditionally at every graph layer.

Existing work therefore focuses on reducing the cost of a single graph
construction or graph convolution pass---whether through window
partitioning \cite{Spadaro_2025_WACV}, dedicated $k$-NN graph-construction
pipelines \cite{Liu_2023_FNNG,He_2024_LFNNG,Neu_2024_GraphGen}, or adaptive
FPGA graph convolution \cite{Tian_2022_FP_GNN,Que_2024_LL_GNN}---whereas
the SpecReuse algorithm targets the complementary architectural problem of reducing the
frequency with which the DIGC stage executes, realizing graph
reuse through a lightweight prediction front-end that uses spectral
similarity to determine whether the graph produced by a previous DIGC
execution can be safely reused
\cite{chung1997spectral,belkin2003laplacian,luxburg2007spectral,wilson2008spectra,munir2025searchvig}. Because the DIGC implementation
remains unchanged, the SpecReuse accelerator is orthogonal to existing FPGA
graph-construction accelerators and, through algorithm--architecture co-design, introduces graph reuse as a new
architectural optimization dimension for FPGA Vision GNN accelerators.

%%%%%%%%%%%%%%%%%%%%%%%%%%%%%%%%%%%%%%%%%%%%%%%%%%%%%%%%%%%%%%%%%%%%%%%%%%%%%%%%
\section{Preliminaries}
\label{sec:preliminaries}

Vision Graph Neural Networks (ViGs) represent an image as a graph whose
vertices correspond to image patches and whose edges encode feature-space
relationships between patches
\cite{VisionGNN,han2023vihgnn,wu2023pvg}. Unlike conventional convolutional
networks with fixed receptive fields, ViGs dynamically update graph
connectivity throughout inference to reflect the evolving semantic
representation of the input. As node features become increasingly
discriminative across successive graph layers, the underlying graph topology
is repeatedly reconstructed, enabling message passing to adapt to the current
feature space.

At graph layer $l$, message passing is expressed as

\begin{equation}
\mathbf{x}_i^{(l)}
=
\Psi^{(l)}
\left(
\mathbf{x}_i^{(l-1)},
\bigoplus_{j\in\mathcal{N}(i)}
\Phi^{(l)}
\left(
\mathbf{x}_i^{(l-1)},
\mathbf{x}_j^{(l-1)}
\right)
\right),
\label{eq:gnn}
\end{equation}

where $\Phi^{(l)}(\cdot)$ and $\Psi^{(l)}(\cdot)$ denote the learnable message and update
functions, respectively, $\oplus$ represents neighborhood aggregation, and
$\mathcal{N}(i)$ denotes the dynamically constructed neighborhood of node
$i$. Consequently, the quality of the graph constructed before each message
message-passing layer directly influences the effectiveness of information
propagation throughout the network.

%%%%%%%%%%%%%%%%%%%%%%%%%%%%%%%%%%%%%%%%%%%%%%%%%%%%%%%%%%%%%%%%%%%%%%%%%%%%%%%%
\subsection{Dynamic Image Graph Construction}
\label{sec:digc}

Before every graph convolution layer, Vision GNNs execute the Dynamic Image
Graph Construction (DIGC) stage to regenerate graph connectivity from the
current node feature representations
\cite{VisionGNN,wang2019dgcnn}. Given the updated feature embeddings, DIGC
computes feature-space distances between nodes, incorporates positional
information, identifies the top-$k$ nearest neighbors, and materializes the
resulting graph adjacency used by the subsequent message passing layer.
Pyramid ViG architectures further reduce the computational complexity of this
process by constructing graphs from a reduced set of co-node features during
early network stages
\cite{wu2023pvg}.

Although DIGC is fundamental to the adaptive behavior of Vision GNNs, it is
executed before every graph layer because node representations continuously
evolve throughout inference. Repeated pairwise distance computation,
neighbor ranking, and graph materialization introduce irregular memory
accesses and comparison-intensive computation, making graph construction a
recurring source of computational cost in Vision GNN inference
\cite{wang2019dgcnn,li2019deepgcns}. Rather than modifying the DIGC procedure
itself, the SpecReuse algorithm addresses the complementary problem of reducing how often
graph construction must be performed. The following section introduces the
SpecReuse algorithm, which exploits the observation that graph connectivity
often changes only gradually between adjacent graph layers.

%%%%%%%%%%%%%%%%%%%%%%%%%%%%%%%%%%%%%%%%%%%%%%%%%%%%%%%%%%%%%%%%%%%%%%%%%%%%%%%%
\section{SpecReuse Architecture}
\label{sec:architecture}

SpecReuse augments the baseline Vision GNN accelerator with a
lightweight prediction front-end that determines whether the Dynamic Image
Graph Construction (DIGC) stage needs to execute. Rather than
reconstructing the graph unconditionally at every graph convolution layer,
the SpecReuse algorithm estimates the structural similarity between the current feature
representation and the previously reconstructed graph using compact spectral
descriptors extracted by a lightweight sparse affinity operator. This
operator is purpose-built for descriptor extraction and is
resource-efficient relative to DIGC's dynamic $k$-NN construction---it performs no pairwise
distance ranking, sorting, or neighbor selection---so it serves only as a
low-overhead proxy signal rather than a partial re-implementation of dynamic graph
reconstruction. If the measured spectral drift remains below a model-specific
reuse threshold, the previously reconstructed graph is reused and the entire
DIGC stage is bypassed; otherwise, the baseline DIGC pipeline executes to
reconstruct a new graph. By modifying only the graph reconstruction policy
rather than the graph-construction algorithm itself, the SpecReuse algorithm
can be integrated with existing FPGA graph-construction accelerators without
altering their underlying implementation or graph convolution pipeline.

The reuse threshold $\tau$ and descriptor dimension $K$ are both
determined offline during model characterization and remain fixed throughout
inference. For each trained ViG backbone, the SpecReuse algorithm evaluates the
distribution of spectral descriptor drift on a validation set and selects
the largest $\tau$, at a fixed $K$, that preserves the target prediction
accuracy while maximizing graph reuse; $K$ is set via the same offline sweep
to the smallest value at which further increases yield no measurable
improvement in reuse-decision accuracy. Since both parameters are computed
once during deployment, the runtime reuse decision reduces to a simple
comparison between the measured descriptor drift and the precomputed
threshold, eliminating the need for adaptive threshold tuning or runtime
calibration.

%%%%%%%%%%%%%%%%%%%%%%%%%%%%%%%%%%%%%%%%%%%%%%%%%%%%%%%%%%%%%%%%%%%%%%%%%%%%%%%%

\subsection{Accelerator Design Motivation}
\label{sec:motivation}

The SpecReuse algorithm reduces DIGC activity only if the reuse decision
itself remains cheap relative to the DIGC execution it replaces. The
affinity operator underlying descriptor extraction is lightweight relative
to DIGC's dynamic $k$-NN construction, but this algorithmic advantage is not
automatically preserved by a software or general-purpose implementation of Algorithm~\ref{alg:specreuse}. Extracting the
spectral descriptor still requires applying Lanczos iteration to obtain the
affinity graph's dominant eigenvalues; on a general-purpose core or a naive RTL
datapath, the repeated sparse matrix-vector products, irregular memory
accesses, and vector reduction operations inherent to Lanczos iteration can
erode a significant fraction of the algorithmic savings, even though the
underlying affinity operator itself is lightweight. A conventional implementation
therefore risks leaving substantial performance on the table: the prediction
front-end that determines whether to skip DIGC would not realize its full
algorithmic advantage. The SpecReuse accelerator eliminates this residual
overhead by mapping Lanczos-based descriptor extraction and drift comparison
onto specialized, fixed-latency hardware, preserving the exact reuse
decisions of Algorithm~\ref{alg:specreuse} while ensuring the front-end's
hardware cost tracks its algorithmic cost. The remainder of this section
describes this accelerator.

\subsection{Architecture Overview}
\label{sec:overview}

\begin{figure*}[!t]\centering \includegraphics[width=0.85\textwidth]{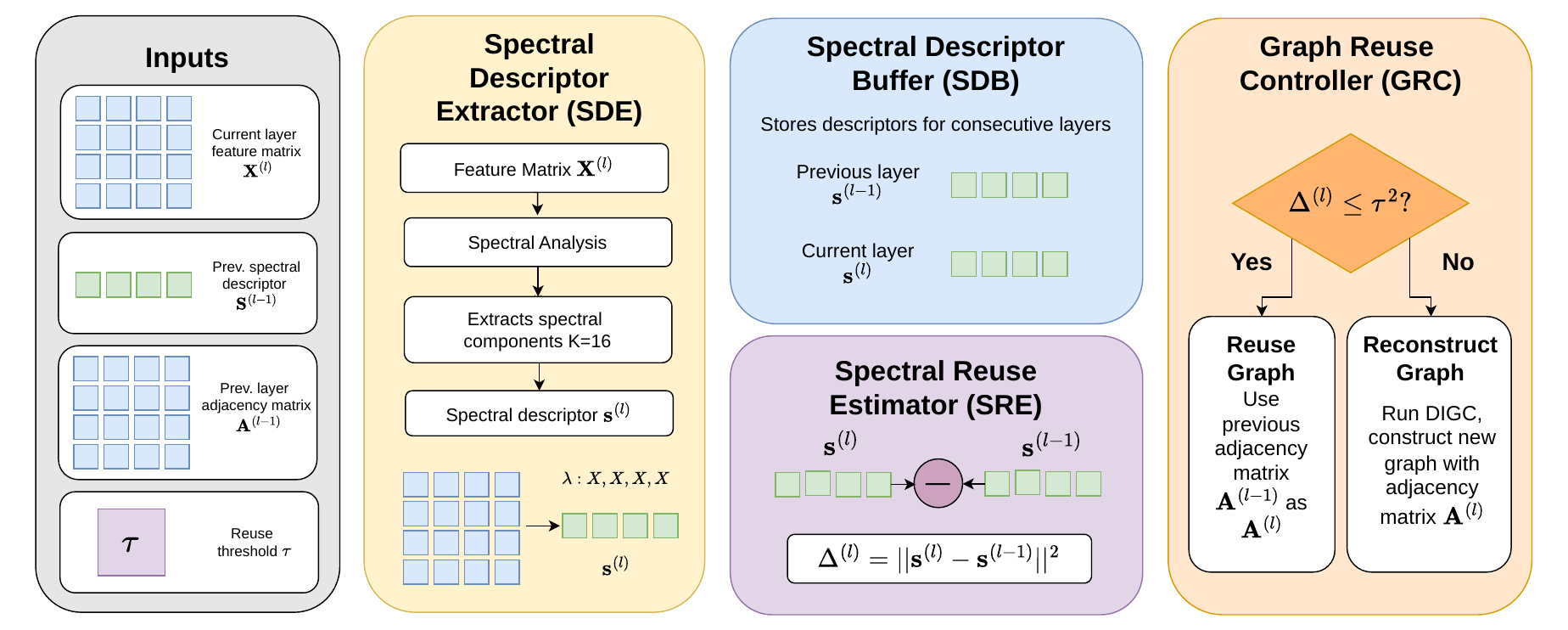} \caption{SpecReuse augments the baseline Dynamic Image Graph Construction (DIGC) execution with a lightweight prediction front-end positioned ahead of the DIGC stage. The front-end predicts whether the graph generated by the previous DIGC execution can be reused before the DIGC stage begins. When graph reuse is selected, the stored adjacency bypasses the DIGC stage and is forwarded directly to graph convolution; otherwise, the baseline DIGC pipeline executes unchanged.} \label{fig:architecture} \vspace{-2mm}\end{figure*}

Figure~\ref{fig:architecture} illustrates the overall SpecReuse architecture.
The prediction front-end is inserted immediately before the DIGC stage,
allowing the graph reuse decision to be made before graph reconstruction
begins. Incoming feature representations are processed by the Spectral
Descriptor Extractor (SDE), which applies a lightweight sparse affinity
operator---distinct from, and resource-efficient relative to, DIGC's dynamic
$k$-NN construction---and computes a compact spectral descriptor from it
using the Lanczos algorithm. The current descriptor is forwarded to the
Spectral Reuse Estimator (SRE), while the Spectral Descriptor Buffer (SDB)
supplies the descriptor of the most recently reconstructed graph. The SRE
computes the drift between these descriptors according to Eq.~(\ref{eq:drift}). The Graph Reuse
Controller (GRC) then compares the measured drift against the
offline-calibrated reuse threshold $\tau$ to determine whether graph
reconstruction is required.

When graph reuse is selected, the controller bypasses the DIGC stage
entirely and retrieves the cached graph adjacency from the graph buffer,
implemented by the Adjacency Register File (ARF). The cached graph adjacency
is then forwarded directly to the graph convolution engine. Otherwise, the
baseline DIGC pipeline executes to reconstruct a new graph from the current
feature representation. Upon completion, the newly reconstructed graph
adjacency is written back to the graph buffer, while its corresponding
spectral descriptor is written to the descriptor buffer. These updated
values become the reference for subsequent graph reuse decisions.

A key property of the SpecReuse accelerator is that graph convolution observes an identical
interface regardless of the selected execution path. Every graph convolution
layer receives a valid graph adjacency through the same input interface,
whether it originates from the graph buffer or from a new DIGC execution.
Consequently, the SpecReuse accelerator requires no modifications to the
downstream graph convolution pipeline and preserves compatibility with
existing DIGC implementations.

The prediction front-end consists of five hardware modules: the Spectral
Descriptor Extractor (SDE), Spectral Descriptor Buffer (SDB), Spectral Reuse
Estimator (SRE), Graph Reuse Controller (GRC), and Adjacency Register File
(ARF). Each module maps directly to a step of Algorithm~\ref{alg:specreuse}:
the SDE realizes affinity construction and descriptor extraction (lines
1--3), the SRE realizes drift computation (line 4), and the GRC realizes the
reuse decision (lines 5--10), with the SDB and ARF providing the persistent
state $s^{(p)}$ and $A^{(p)}$ the decision depends on. Together, these
modules implement graph reuse entirely ahead of the DIGC stage while
preserving the baseline graph construction and graph convolution datapaths.
The following subsections describe the design and operation of each module
in detail.
%%%%%%%%%%%%%%%%%%%%%%%%%%%%%%%%%%%%%%%%%%%%%%%%%%%%%%%%%%%%%%%%%%%%%%%%%%%%%%%%
\subsection{Spectral Descriptor-Based Graph Reuse}
\label{sec:similarity}

Dynamic graph construction is one of the primary computational bottlenecks in
Vision Graph Neural Networks (ViGs), as each graph convolution layer
reconstructs a dynamic $k$-nearest-neighbor (k-NN) graph from the current node
features. Although node embeddings evolve throughout the network, the induced
graph structure often changes gradually between consecutive layers,
particularly in deeper stages where the feature representations have largely
converged. Consequently, reconstructing the graph at every layer frequently
performs redundant computation while producing highly similar graphs.

The SpecReuse algorithm exploits this observation by predicting whether the graph that would
be constructed from the current feature representation differs sufficiently
from the previously reconstructed graph to justify reconstruction. Rather than
constructing successive graphs for direct comparison, it compares
compact spectral descriptors extracted from a lightweight sparse affinity
graph computed specifically for this purpose, independent of DIGC's dynamic
$k$-NN construction. Spectral graph representations have been widely used for graph comparison, graph matching, and structure-preserving embedding because they capture global structural properties in a compact representation while remaining invariant to node ordering
\cite{chung1997spectral,belkin2003laplacian,luxburg2007spectral,wilson2008spectra,munir2025searchvig}. The algorithm
leverages these spectral representations as a proxy for structural similarity, allowing dynamic graph reconstruction to be avoided entirely whenever the graph remains sufficiently stable.

Let $X^{(l)} \in \mathbb{R}^{N\times d}$ denote the node feature matrix at graph convolution layer $l$, where $N$ is
the number of graph nodes and $d$ is the feature dimension. SpecReuse first
computes a lightweight sparse affinity graph, dedicated to descriptor
extraction and independent of DIGC's own dynamic $k$-NN construction, $G^{(l)} = \mathcal{S}_{\mathrm{ViG}}\!\left(X^{(l)}\right)$, where $\mathcal{S}_{\mathrm{ViG}}(\cdot)$ denotes this lightweight affinity
operator, purpose-built for spectral descriptor extraction and used
exclusively to inform the graph reuse decision. Unlike DIGC's pairwise
distance evaluation, top-$k$ ranking, and adjacency materialization,
$\mathcal{S}_{\mathrm{ViG}}(\cdot)$ produces only a sparse structural summary
sufficient for spectral descriptor extraction, at lower resource cost than
full dynamic graph reconstruction.

Rather than constructing the complete eigenspectrum of $G^{(l)}$, the SpecReuse algorithm
computes only its $K$ dominant eigenvalues, where $K \ll N$ is a fixed
descriptor dimension chosen offline, using the Lanczos algorithm,

\begin{equation}
\left\{
\lambda_1^{(l)},
\lambda_2^{(l)},
\ldots,
\lambda_K^{(l)}
\right\}
=
\operatorname{Lanczos}
\!\left(
G^{(l)},K
\right),
\label{eq:lanczos}
\end{equation}

where the eigenvalues are ordered in descending order,

\[
\lambda_1^{(l)}
\ge
\lambda_2^{(l)}
\ge
\cdots
\ge
\lambda_K^{(l)}.
\]

The spectral descriptor associated with layer $l$ is defined as

\begin{equation}
s^{(l)}
=
\left[
\lambda_1^{(l)},
\lambda_2^{(l)},
\ldots,
\lambda_K^{(l)}
\right]^T.
\label{eq:descriptor}
\end{equation}

Retaining only the largest $K$ eigenvalues produces a compact structural
signature that summarizes the dominant connectivity characteristics of the
affinity graph while requiring substantially less computation and storage than
the complete graph representation.

Let $s^{(p)}$ denote the spectral descriptor associated with the most recently
reconstructed graph and let $A^{(p)}$ denote the corresponding adjacency
matrix. Structural change between consecutive layers is quantified by the
descriptor drift

\begin{equation}
\Delta^{(l)}
=
\left\|
s^{(l)}
-
s^{(p)}
\right\|_2^2,
\label{eq:drift}
\end{equation}

which measures the distance between the current and previously reconstructed
spectral descriptors. The squared Euclidean distance preserves the same
ordering as the Euclidean norm while eliminating the square-root operation.

Graph reconstruction is performed only when the descriptor drift exceeds a
programmable threshold,

\begin{equation}
A^{(l)}
=
\begin{cases}
A^{(p)},
&
\Delta^{(l)}
\le
\tau^2,
\\[6pt]
\operatorname{kNN}(X^{(l)},k),
&
\Delta^{(l)}
>
\tau^2,
\end{cases}
\label{eq:reuse}
\end{equation}

where $\tau$ controls the trade-off between graph reuse and graph
reconstruction. When the descriptor drift remains below the threshold, the
current graph is considered spectrally similar to the previously
reconstructed graph, and the existing graph is reused. Otherwise, a new graph
is reconstructed from the current feature representation and its spectral
descriptor replaces the previously stored descriptor.

Algorithm~\ref{alg:specreuse} summarizes the complete graph reuse procedure.
Beginning with the current feature representation, the SpecReuse algorithm constructs a
lightweight affinity graph using an operator dedicated to descriptor
extraction, extracts a compact spectral descriptor using the Lanczos
algorithm, measures descriptor drift relative to the previously reconstructed
graph, and determines whether dynamic graph reconstruction is necessary. Because this
decision is made before dynamic graph construction and relies only on this
lightweight proxy, the full DIGC pipeline---pairwise distance computation,
top-$k$ selection, and adjacency materialization---can be eliminated
entirely whenever the graph remains spectrally stable.

\begin{algorithm}[t]
\caption{SpecReuse Algorithm: Spectral Descriptor-Based Graph Reuse}
\label{alg:specreuse}
\footnotesize
\begin{algorithmic}[1]

\Require Current feature matrix $X^{(l)}$, previous spectral descriptor $s^{(p)}$, previous graph adjacency $A^{(p)}$, graph degree $k$, descriptor dimension $K$, reuse threshold $\tau$

\Ensure Graph adjacency $A^{(l)}$

\State Construct the sparse ViG affinity graph:
$G^{(l)} \gets \mathcal{S}_{\mathrm{ViG}}(X^{(l)})$

\State Compute the largest $K$ eigenvalues using Lanczos:
$\{\lambda_1^{(l)},\ldots,\lambda_K^{(l)}\} \gets \operatorname{Lanczos}(G^{(l)},K)$

\State Form the spectral descriptor:
$s^{(l)} \gets [\lambda_1^{(l)},\ldots,\lambda_K^{(l)}]^T$

\State Compute descriptor drift:
$\Delta^{(l)} \gets \|s^{(l)}-s^{(p)}\|_2^2$

\If{$\Delta^{(l)}\le\tau^2$}
    \State Reuse the previous graph:
    $A^{(l)} \gets A^{(p)}$
\Else
    \State Reconstruct the graph:
    $A^{(l)} \gets \operatorname{kNN}(X^{(l)},k)$
    \State Update stored descriptor:
    $s^{(p)} \gets s^{(l)}$
    \State Update stored graph:
    $A^{(p)} \gets A^{(l)}$
\EndIf

\State \Return $A^{(l)}$

\end{algorithmic}
\end{algorithm}
%%%%%%%%%%%%%%%%%%%%%%%%%%%%%%%%%%%%%%%%%%%%%%%%%%%%%%%%%%%%%%%%%%%%%%%%%%%%%%%%
\subsection{FPGA Integration}
\label{sec:integration}

The SpecReuse accelerator is integrated as a lightweight prediction front-end
that precedes the DIGC pipeline in its entirety, using its own dedicated
affinity operator---distinct from, and resource-efficient relative to, DIGC's dynamic $k$-NN
construction---to decide whether DIGC executes. This organization preserves
compatibility with the baseline ViG pipeline and requires no modification to
the graph convolution datapath, which receives a valid adjacency through the
same interface whether it originates from the graph buffer or a new DIGC
execution (\S\ref{sec:overview}).

The hardware overhead of this integration is modest because the SDE, SRE,
and GRC operate on a lightweight affinity graph and a $K$-dimensional
descriptor rather than on the dense candidate neighborhoods DIGC evaluates.
The Spectral Reuse Estimator computes Eq.~(\ref{eq:drift}) through
fixed-length element-wise subtraction, squaring, and accumulation over the
retained eigenvalues, while the Graph Reuse Controller implements a single
threshold comparison to select between the reuse and reconstruction paths.
On-chip storage is limited to the graph buffer (ARF) and descriptor buffer
(SDB), which hold one cached adjacency and one $K$-element descriptor
between reconstruction events. Because $K$ is fixed and independent of graph
connectivity, and the affinity operator performs no pairwise ranking, sorting, or
neighbor selection, the front-end's cost does not scale with DIGC's dominant
cost drivers, so its overhead remains small relative to a DIGC invocation
and is fully amortized whenever that invocation is eliminated.
%%%%%%%%%%%%%%%%%%%%%%%%%%%%%%%%%%%%%%%%%%%%%%%%%%%%%%%%%%%%%%%%%%%%%%%%%%%%%%%%

%%%%%%%%%%%%%%%%%%%%%%%%%%%%%%%%%%%%%%%%%%%%%%%%%%%%%%%%%%%%%%%%%%%%%%%%%%%%%%%%

\section{Evaluation}
\label{sec:evaluation}

We evaluate SpecReuse on a set of ViG
backbones to answer four questions: (1) what FPGA overhead does SpecReuse
introduce, (2) how much graph reconstruction does it eliminate, (3) how much
end-to-end latency reduction does graph reuse provide, and (4) how does the
graph reuse threshold affect the trade-off between prediction accuracy and graph
reconstruction reduction.

%%%%%%%%%%%%%%%%%%%%%%%%%%%%%%%%%%%%%%%%%%%%%%%%%%%%%%%%%%%%%%%%%%%%%%%%%%%%%%%%
\subsection{Experimental Methodology}

All experiments compare the baseline Dynamic Image Graph Construction (DIGC)
accelerator against the same accelerator augmented with SpecReuse. CPU and GPU
implementations are included as software reference platforms, while FPGA
results characterize the architectural impact of graph reuse and are implemented using High-Level Synthesis (HLS). All Vision GNN
models are evaluated on the ImageNet-1K validation set using the standard
single-crop evaluation protocol
\cite{VisionGNN,greedyvig2024,clustervig2025,imagenet2015}. Classification
accuracy is measured using a batch size of 128, while latency measurements use
a batch size of one to reflect deployment-oriented inference performance. For each Vision GNN backbone, the graph reuse threshold $\tau$ is calibrated
offline prior to deployment. Threshold calibration sweeps $\tau$ across the
validation set while measuring both classification accuracy and graph reuse.
The selected operating point corresponds to the threshold that provides the
highest graph reuse while maintaining the desired prediction accuracy. Once
selected, $\tau$ remains fixed throughout inference and is applied uniformly to
all graph reuse decisions. Table~\ref{tab:setup} summarizes the experimental platform.

\begin{table}
\centering
\footnotesize
\setlength{\tabcolsep}{6pt}
\renewcommand{\arraystretch}{0.95}
\caption{Experimental Platform}
\label{tab:setup}
\begin{tabular}{ll}
\toprule
\textbf{Hardware} & \textbf{Configuration} \\
\midrule
FPGA & AMD Alveo U280 @ 300 MHz \\
Toolchain & AMD Vivado/Vitis 2024.1 \\
CPU & AMD EPYC 7313 (16-core) \\
GPU & NVIDIA RTX 6000 Ada \\
Dataset & ImageNet-1K \\
Precision & FP16 \\
Batch Size & 128 (accuracy), 1 (latency) \\
\bottomrule
\end{tabular}
\end{table}

%%%%%%%%%%%%%%%%%%%%%%%%%%%%%%%%%%%%%%%%%%%%%%%%%%%%%%%%%%%%%%%%%%%%%%%%%%%%%%%%
\subsection{FPGA Implementation Overhead}

We first evaluate the hardware cost of integrating SpecReuse into the baseline
accelerator. Table~\ref{tab:resources} reports the incremental FPGA resources
required by the spectral descriptor generation, similarity evaluation, reuse
control, and graph selection logic. SpecReuse introduces only modest
additional logic and on-chip memory while preserving the baseline operating
frequency of 300\,MHz, demonstrating that graph reuse can be incorporated with
minimal hardware overhead. The additional hardware increases LUT, flip-flop,
DSP, and BRAM utilization by only $1.07\%$, $0.56\%$, $0.12\%$, and $0.88\%$,
respectively, relative to the baseline accelerator, indicating that the graph reuse mechanism scales
efficiently with the existing accelerator datapath.

\begin{table}
\centering
\footnotesize
\setlength{\tabcolsep}{4pt}
\renewcommand{\arraystretch}{0.95}
\caption{FPGA resource utilization. The \emph{SpecReuse} row reports only the additional hardware introduced by the spectral reuse components (SDE, SDB, SRE, GRC, and ARF). Both the baseline accelerator and SpecReuse operate at 300\,MHz.}
\label{tab:resources}
\begin{tabular}{lccccc}
\toprule
\textbf{Design} &
\textbf{LUT} &
\textbf{FF} &
\textbf{DSP} &
\textbf{BRAM} &
\textbf{URAM} \\
\midrule
Baseline  & 650,232 & 1,173,312 & 5,866 & 1,250 & 768 \\
SpecReuse & 6,925   & 6,512     & 7     & 11    & 0 \\
\midrule
Incremental & 1.07\% & 0.56\% & 0.12\% & 0.88\% & 0.00\% \\
\bottomrule
\end{tabular}
\end{table}

%%%%%%%%%%%%%%%%%%%%%%%%%%%%%%%%%%%%%%%%%%%%%%%%%%%%%%%%%%%%%%%%%%%%%%%%%%%%%%%%

\begin{table}
\centering
\footnotesize
\renewcommand{\arraystretch}{0.95}
\setlength{\tabcolsep}{4pt}
\caption{End-to-end inference latency comparison across CPU, GPU, the baseline FPGA accelerator, and the proposed SpecReuse-enhanced FPGA accelerator. SpecReuse achieves up to $2.69\times$ speedup over the baseline FPGA implementation.}
\label{tab:performance}
\begin{tabular}{lccccc}
\toprule
\textbf{Model} &
\textbf{CPU} &
\textbf{GPU} &
\textbf{Baseline} &
\textbf{SpecReuse} &
\textbf{Speedup} \\
&
(ms) &
(ms) &
(ms) &
(ms) &
($\times$) \\
\midrule
ViG-Ti  & 67.8  & 13.4 & 2.88 & 1.07 & $2.69\times$ \\
ViG-S   & 109.4 & 16.2 & 2.94 & 1.13 & $2.60\times$ \\
ViG-B   & 224.4 & 16.7 & 4.16 & 1.72 & $2.42\times$ \\
PViG-Ti & 89.8  & 6.41 & 2.88 & 1.16 & $2.48\times$ \\
PViG-S  & 118.8 & 4.79 & 4.68 & 1.93 & $2.42\times$ \\
PViG-M  & 177.3 & 6.33 & 5.85 & 2.48 & $2.36\times$ \\
PViG-B  & 168.2 & 5.22 & 6.66 & 2.96 & $2.25\times$ \\
\bottomrule
\end{tabular}
\end{table}

\begin{table}
\centering
\footnotesize
\renewcommand{\arraystretch}{0.95}
\setlength{\tabcolsep}{5pt}
\caption{Average energy per inference across GPU and FPGA implementations. FPGA energy is computed as $E=P\times t$ using post-place-and-route Vivado power estimates and measured inference latency.}
\label{tab:energy}
\resizebox{0.48\textwidth}{!}{
\begin{tabular}{lccccccc}
\toprule
 & \textbf{ViG-Ti} & \textbf{ViG-S} & \textbf{ViG-B} &
 \textbf{PViG-Ti} & \textbf{PViG-S} & \textbf{PViG-M} & \textbf{PViG-B} \\
\midrule
\textbf{GPU (mJ)}
& 3733.7 & 4399.0 & 4772.4 & 876.8 & 737.8 & 948.5 & 939.9 \\

\textbf{Baseline FPGA (mJ)}
& 60.5 & 61.8 & 87.4 & 60.5 & 98.3 & 122.9 & 139.9 \\

\textbf{SpecReuse FPGA (mJ)}
& 21.2 & 22.4 & 34.1 & 23.0 & 38.3 & 49.2 & 58.7 \\
\bottomrule
\end{tabular}
}
\end{table}

\subsection{End-to-End Performance}

\begin{figure}
\centering
\includegraphics[width=\columnwidth]{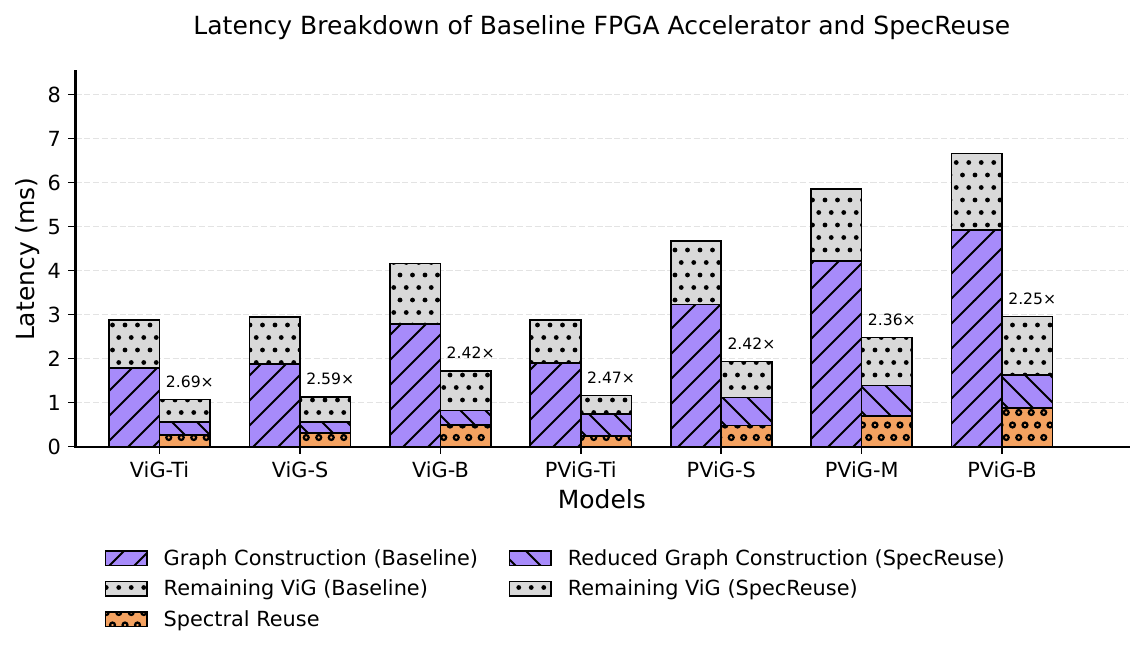}
\caption{Inference latency breakdown of baseline accelerator vs. SpecReuse.}
\label{fig:latency}
\vspace{-2mm}
\end{figure}

Table~\ref{tab:performance} shows that SpecReuse achieves
$2.25\times$--$2.69\times$ speedup over the baseline FPGA accelerator and
outperforms the CPU and GPU implementations. As shown in
Fig.~\ref{fig:latency}, these gains primarily result from eliminating redundant
Dynamic Image Graph Construction (DIGC), while the remaining network latency is
largely unchanged. Table~\ref{tab:energy} reports average energy per inference,
with FPGA energy computed as $E=P\times t$ from post-place-and-route power
estimates and measured latency. By shortening inference with modest hardware
overhead, SpecReuse reduces FPGA energy by approximately $58\%$--$65\%$ across
the evaluated Vision GNN backbones.

%%%%%%%%%%%%%%%%%%%%%%%%%%%%%%%%%%%%%%%%%%%%%%%%%%%%%%%%%%%%%%%%%%%%%%%%%%%%%%%%
\subsection{Reuse Threshold Sensitivity}
The graph reuse threshold $\tau$ controls the trade-off between graph reuse and classification accuracy. Larger thresholds increase graph reuse by eliminating more Dynamic Image Graph Construction (DIGC) invocations, but eventually introduce prediction error as outdated graph connectivity is reused. For each Vision GNN backbone, $\tau$ is calibrated offline by sweeping candidate thresholds on the ImageNet validation set and measuring both Top-1 accuracy and graph reuse. Table~\ref{tab:tau} summarizes the selected threshold, graph reuse, and corresponding classification accuracy for each model. The calibrated thresholds demonstrate that substantial graph reuse can be achieved with only minor reductions in prediction accuracy. Across all evaluated Vision GNN backbones, the selected operating points enable $18.1\%$--$30.1\%$ graph reuse while limiting Top-1 accuracy degradation to at most $0.54$ percentage points. Larger Pyramid ViG variants generally tolerate higher reuse rates than standard ViG models, indicating greater structural similarity between successive graph layers. These calibrated thresholds are used for all reported performance and energy evaluations.

\begin{table}
\centering
\footnotesize
\setlength{\tabcolsep}{5pt}
\renewcommand{\arraystretch}{0.95}
\caption{Offline threshold calibration. For each Vision GNN backbone, the graph reuse threshold $\tau$ is calibrated offline by selecting the operating point that maximizes graph reuse while limiting accuracy degradation.}
\label{tab:tau}
\resizebox{0.48\textwidth}{!}{
\begin{tabular}{lccccc}
\toprule
\textbf{Model} &
\textbf{Selected $\tau$} &
\textbf{Reuse (\%)} &
\textbf{Baseline Top-1 (\%)} &
\textbf{Calibrated Top-1 (\%)} &
\textbf{Accuracy Loss (pp)} \\
\midrule
ViG-Ti  & 2.1549 & 18.1 & 73.90 & 73.68 & 0.22 \\
ViG-S   & 2.4344 & 22.4 & 80.40 & 80.13 & 0.27 \\
ViG-B   & 1.9827 & 28.8 & 82.30 & 81.79 & 0.51 \\
PViG-Ti & 2.5183 & 26.9 & 78.50 & 78.17 & 0.33 \\
PViG-S  & 2.5464 & 29.1 & 82.10 & 81.85 & 0.25 \\
PViG-M  & 2.5695 & 25.4 & 83.10 & 82.82 & 0.28 \\
PViG-B  & 2.9217 & 30.1 & 83.70 & 83.16 & 0.54 \\
\bottomrule
\end{tabular}}
\end{table}

%%%%%%%%%%%%%%%%%%%%%%%%%%%%%%%%%%%%%%%%%%%%%%%%%%%%%%%%%%%%%%%%%%%%%%%%%%%%%%%%
\subsection{Evaluation Summary}

Across all evaluated ViG backbones, SpecReuse consistently reduces
Dynamic Image Graph Construction activity with minimal FPGA implementation
overhead. The resulting reduction in graph construction translates directly
into lower end-to-end inference latency and lower energy per inference while
preserving classification accuracy.
Overall, these results demonstrate that graph reuse is an effective
architectural optimization for FPGA Vision GNN accelerators, enabling
substantial performance and energy improvements without modifying the
underlying ViG architecture.

\section{Conclusion}

This paper presented SpecReuse, an FPGA accelerator that uses compact spectral descriptors to selectively reuse graphs and reduce Dynamic Image Graph Construction (DIGC) overhead in ViGs. Across seven ViG and Pyramid ViG backbones, SpecReuse adds only \mbox{$1.07\%$} LUT, \mbox{$0.56\%$} FF, \mbox{$0.12\%$} DSP, and \mbox{$0.88\%$} BRAM overhead while achieving up to a \mbox{$2.69\times$} end-to-end inference speedup and approximately \mbox{$58$--$65\%$} lower energy per inference. Offline threshold calibration enables \mbox{$18.1$--$30.1\%$} graph reuse while limiting Top-1 accuracy degradation to at most \mbox{$0.54$} percentage points. Future works will include exploring the impact of expanding the spectral descriptor buffer and comparing other types of threshold similarity with spectral similarity.

\section{Acknowledgements}
This work was supported in part by the ARO grants W911NF-242-0194 and and by the National Science Foundation grants OAC-2411446 and OAC-2505107.

%\clearpage
\bibliographystyle{IEEEtran}
\bibliography{IEEEabrv,citation}

\end{document}